\documentclass[reprint,amsmath,amssymb,aps,prd]{revtex4-1}
\usepackage{bm}
\usepackage{dcolumn}
\usepackage{lineno}
\usepackage{graphicx}
\usepackage{subfigure}
\usepackage{epstopdf}
\usepackage{float}
\usepackage{threeparttable}
\usepackage{overpic}
\usepackage{subfigure}
\usepackage{float}
\usepackage{enumitem}
\usepackage{bm}
\usepackage{cmap} 
\usepackage{comment}
\usepackage{mathtext}
\usepackage{mathrsfs}
\usepackage{multirow}
\usepackage{xcolor}
\definecolor{aa}{RGB}{0,0,139}
\usepackage[breaklinks]{hyperref}
\hypersetup{bookmarksnumbered=true,colorlinks,urlcolor=aa,linkcolor=blue,anchorcolor=aa,citecolor=aa}
\begin{document}

\newcommand{\afs}{\alpha_s}
\newcommand{\bgp}{\beta\gamma}
\newcommand{\eff}{\varepsilon}
\newcommand{\sintht}{\sin{\theta}}
\newcommand{\costht}{\cos{\theta}}
\newcommand{\dedx}{dE/dx}

\newcommand{\probfc}{Prob_{\chi^2}}
\newcommand{\probpi}{Prob_{\pi}}
\newcommand{\probka}{Prob_{K}}
\newcommand{\probpr}{Prob_{p}}
\newcommand{\proball}{Prob_{all}}

\newcommand{\chicJ}{\chi_{cJ}}
\newcommand{\gchicJ}{\gamma\chi_{cJ}}
\newcommand{\gchica}{\gamma\chi_{c0}}
\newcommand{\gchicb}{\gamma\chi_{c1}}
\newcommand{\gchicc}{\gamma\chi_{c2}}
\newcommand{\hc}{h_c(^1p_1)}
\newcommand{\qqb}{q\bar{q}}
\newcommand{\uub}{u\bar{u}}
\newcommand{\ddb}{d\bar{d}}
\newcommand{\SSB}{\Sigma^+\bar{\Sigma}^-}
\newcommand{\ccb}{c\bar{c}}

\newcommand{\psipto}{\psi^{\prime}\rightarrow \pi^+\pi^- J/\psi}
\newcommand{\ptomm}{J/\psi\rightarrow \mu^+\mu^-}
\newcommand{\ppp}{\pi^+\pi^- \pi^0}
\newcommand{\pip}{\pi^+}
\newcommand{\pim}{\pi^-}
\newcommand{\kap}{K^+}
\newcommand{\kam}{K^-}
\newcommand{\ks}{K^0_s}
\newcommand{\pbar}{\bar{p}}
\newcommand{\jp}{J/\psi\rightarrow \gamma\pi^0}
\newcommand{\je}{J/\psi\rightarrow \gamma\eta}
\newcommand{\jep}{J/\psi\rightarrow \gamma\eta^{\prime}}

\newcommand{\LL}{\ell^+\ell^-}
\newcommand{\EE}{e^+e^-}
\newcommand{\MM}{\mu^+\mu^-}
\newcommand{\GG}{\gamma\gamma}
\newcommand{\TT}{\tau^+\tau^-}
\newcommand{\pp}{\pi^+\pi^-}
\newcommand{\kk}{K^+K^-}
\newcommand{\ppb}{p\bar{p}}
\newcommand{\gpp}{\gamma \pi^+\pi^-}
\newcommand{\gkk}{\gamma K^+K^-}
\newcommand{\gppb}{\gamma p\bar{p}}
\newcommand{\ggee}{\gamma\gamma e^+e^-}
\newcommand{\gguu}{\gamma\gamma\mu^+\mu^-}
\newcommand{\ggll}{\gamma\gamma l^+l^-}
\newcommand{\ppee}{\pi^+\pi^- e^+e^-}
\newcommand{\ppuu}{\pi^+\pi^-\mu^+\mu^-}
\newcommand{\etap}{\eta^{\prime}}
\newcommand{\gpi}{\gamma\pi^0}
\newcommand{\geta}{\gamma\eta}
\newcommand{\getap}{\gamma\etap}
\newcommand{\pppp}{\pi^+\pi^-\pi^+\pi^-}
\newcommand{\ppkk}{\pi^+\pi^-K^+K^-}
\newcommand{\pppr}{\pi^+\pi^-p\bar{p}}
\newcommand{\kkkk}{K^+K^-K^+K^-}
\newcommand{\kskp}{K^0_s K^+ \pi^- + c.c.}
\newcommand{\ppkp}{\pi^+\pi^-K^+ \pi^- + c.c.}
\newcommand{\ksks}{K^0_s K^0_s}
\newcommand{\dphi}{\phi\phi}
\newcommand{\phikk}{\phi K^+K^-}
\newcommand{\ppeta}{\pi^+\pi^-\eta}
\newcommand{\gpppp}{\gamma \pi^+\pi^-\pi^+\pi^-}
\newcommand{\gppkk}{\gamma \pi^+\pi^-K^+K^-}
\newcommand{\gpppr}{\gamma \pi^+\pi^-p\bar{p}}
\newcommand{\gkkkk}{\gamma K^+K^-K^+K^-}
\newcommand{\gkskp}{\gamma K^0_s K^+ \pi^- + c.c.}
\newcommand{\gppkp}{\gamma \pi^+\pi^-K^+ \pi^- + c.c.}
\newcommand{\gksks}{\gamma K^0_s K^0_s}
\newcommand{\gphiphi}{\gamma \phi\phi}

\newcommand{\tpp}{3(\pi^+\pi^-)}
\newcommand{\tppkk}{2(\pi^+\pi^-)(K^+K^-)}
\newcommand{\pptkk}{(\pi^+\pi^-)2(K^+K^-)}
\newcommand{\tkk}{3(K^+K^-)}
\newcommand{\gtpp}{\gamma 3(\pi^+\pi^-)}
\newcommand{\gtppkk}{\gamma 2(\pi^+\pi^-)(K^+K^-)}
\newcommand{\gpptkk}{\gamma (\pi^+\pi^-)2(K^+K^-)}
\newcommand{\gtkk}{\gamma 3(K^+K^-)}

\newcommand{\psp}{\psi(3686)}
\newcommand{\jpsi}{J/\psi}
\newcommand{\ar}{\rightarrow}
\newcommand{\lra}{\longrightarrow}
\newcommand{\jpsito}{J/\psi \rightarrow }
\newcommand{\ptoppjp}{J/\psi \rightarrow\pi^+\pi^- J/\psi}
\newcommand{\pspto}{\psi^\prime \rightarrow }
\newcommand{\ptop}{\psi'\rightarrow\pi^0 J/\psi}
\newcommand{\ptoeta}{\psi'\rightarrow\eta J/\psi}
\newcommand{\ecto}{\eta_c \rightarrow }
\newcommand{\ecpto}{\eta_c^\prime \rightarrow }
\newcommand{\xto}{X(3594) \rightarrow }
\newcommand{\chicJto}{\chi_{cJ} \rightarrow }
\newcommand{\chiczto}{\chi_{c0} \rightarrow }
\newcommand{\chicoto}{\chi_{c1} \rightarrow }
\newcommand{\chictto}{\chi_{c2} \rightarrow }
\newcommand{\pspp}{\psi^{\prime\prime}}
\newcommand{\ptochic}{\psi(2S)\ar \gamma\chi_{c1,2}}
\newcommand{\ppjpsi}{\pi^0\pi^0 J/\psi}
\newcommand{\utoeta}{\Upsilon^{\prime}\ar\eta\Upsilon}
\newcommand{\ww}{\omega\omega}
\newcommand{\wf}{\omega\phi}
\newcommand{\ff}{\phi\phi}
\newcommand{\npsp}{N_{\psp}}
\newcommand{\llb}{\Lambda\bar{\Lambda}}
\newcommand{\llbpi}{\llb\pi^0}
\newcommand{\llbeta}{\llb\eta}
\newcommand{\ppi}{p\pi^-}
\newcommand{\pbpi}{\bar{p}\pi^+}
\newcommand{\lamb}{\bar{\Lambda}}
\def\ctup#1{$^{\cite{#1}}$}
\newcommand{\bfg}{\begin{figure}}
\newcommand{\efg}{\end{figure}}
\newcommand{\bitm}{\begin{itemize}}
\newcommand{\eitm}{\end{itemize}}
\newcommand{\bnum}{\begin{enumerate}}
\newcommand{\enum}{\end{enumerate}}
\newcommand{\btbl}{\begin{table}}
\newcommand{\etbl}{\end{table}}
\newcommand{\btbu}{\begin{tabular}}
\newcommand{\etbu}{\end{tabular}}
\newcommand{\bcl}{\begin{center}}
\newcommand{\ecl}{\end{center}}
\newcommand{\bbt}{\bibitem}
\newcommand{\beq}{\begin{equation}}
\newcommand{\eeq}{\end{equation}}
\newcommand{\beqr}{\begin{eqnarray}}
\newcommand{\eeqr}{\end{eqnarray}}
\newcommand{\red}{\color{red}}
\newcommand{\blue}{\color{blue}}
\newcommand{\yellow}{\color{yellow}}
\newcommand{\green}{\color{green}}
\newcommand{\purple}{\color{purple}}
\newcommand{\brown}{\color{brown}}
\newcommand{\black}{\color{black}}

\definecolor{boslv}{rgb}{0.0, 0.65, 0.58}
\definecolor{Munsell}{HTML}{00A877}
\newcommand{\psip}{\psi^{'}}
\newcommand{\psipp}{\psi(3686)}

\newcommand{\Br}{\mathcal{B}}
\newcommand{\too}{\rightarrow}
\newcommand{\del}{\color{red}\sout}
\newcommand{\new}{\color{blue}\uwave}
\title{Observation of the decay \begin{boldmath}$\mathbf{\psi(3686)}\rightarrow\mathbf{\Sigma^{0}\bar{\Sigma}^{0}\omega}$\end{boldmath}}

\author{
\begin{small}
\begin{center}
M.~Ablikim$^{1}$, M.~N.~Achasov$^{4,c}$, P.~Adlarson$^{77}$, X.~C.~Ai$^{82}$, R.~Aliberti$^{36}$, A.~Amoroso$^{76A,76C}$, Q.~An$^{73,59,a}$, Y.~Bai$^{58}$, O.~Bakina$^{37}$, Y.~Ban$^{47,h}$
% ...
\\
\vspace{0.2cm}
(BESIII Collaboration)\\
\vspace{0.2cm} {\it
$^{1}$ Institute of High Energy Physics, Beijing 100049, People's Republic of China\\
% ...
$^{4}$ Budker Institute of Nuclear Physics SB RAS (BINP), Novosibirsk 630090, Russia\\
% ...
$^{36}$ Johannes Gutenberg University of Mainz, Johann-Joachim-Becher-Weg 45, D-55099 Mainz, Germany\\
$^{37}$ Joint Institute for Nuclear Research, 141980 Dubna, Moscow region, Russia\\
% ...
$^{47}$ Peking University, Beijing 100871, People's Republic of China\\
% ...
$^{58}$ Southeast University, Nanjing 211100, People's Republic of China\\
$^{59}$ State Key Laboratory of Particle Detection and Electronics, Beijing 100049, Hefei 230026, People's Republic of China\\
% ...
$^{73}$ University of Science and Technology of China, Hefei 230026, People's Republic of China\\
% ...
$^{76}$ University of Turin and INFN, (A)University of Turin, I-10125, Turin, Italy; (B)University of Eastern Piedmont, I-15121, Alessandria, Italy; (C)INFN, I-10125, Turin, Italy\\
$^{77}$ Uppsala University, Box 516, SE-75120 Uppsala, Sweden\\
% ...
$^{82}$ Zhengzhou University, Zhengzhou 450001, People's Republic of China\\

\vspace{0.2cm}
$^{a}$ Deceased\\
% ...
$^{c}$ Also at the Novosibirsk State University, Novosibirsk, 630090, Russia\\
% ...
$^{h}$ Also at State Key Laboratory of Nuclear Physics and Technology, Peking University, Beijing 100871, People's Republic of China\\
% ...
}
\end{center}
\vspace{0.4cm}
\end{small}
}

\date{March 24, 2025}

	\begin{abstract}
		Using a dataset of $(27.12\pm 0.14)\times 10^{8}$ $\psi(3686)$
                events collected by the BESIII detector operating at
                the BEPCII collider, we report the first observation of the decay
                $\psi(3686)\to\Sigma^{0}\bar{\Sigma}^{0}\omega$ with a statistical
                significance of 8.9$\sigma$. The measured branching fraction is     $(1.24 \pm 0.16_{\textrm{stat}} \pm
                0.11_{\textrm{sys}}) \times 10^{-5}$, where the first
                uncertainty is statistical and the second is
                systematic. Additionally, we investigate potential
                intermediate states in the invariant mass distributions of $\Sigma^{0}\omega$, $\bar{\Sigma}^{0}\omega$ and $\Sigma^{0}\bar{\Sigma}^{0}$. A hint of a resonance is observed in the invariant mass distribution of $M_{\Sigma^{0}(\bar{\Sigma}^{0})\omega}$, located around 2.06 GeV/$c^2$, with a significance of 2.5$\sigma$.
        \end{abstract} 
\maketitle
	
   \section{Introduction}\label{sec:introduction}
  \vspace{-0.3cm}
   
The existing theoretical framework of Quantum Chromodynamics (QCD) successfully describes strong interactions between quarks and gluons in the perturbative regime. However, precise calculations in the non-perturbative regime remain challenging and often depend on effective theories and approximations.
Charmonium decays, which straddle the perturbative and non-perturbative regimes, provide an ideal laboratory for testing QCD theories~\cite{BESIII_physics,QCD_charmonium2}.

In addition to advancing our understanding of charmonium states, studying charmonium decays also facilitate the search for excited baryon resonances. Investigating the hadronic three-body final states of $\psi(3686)$ decays, such as $\psi(3686)\rightarrow B\bar{B}P~(V)$,
where $B$ denotes a baryon, $\bar{B}$ its antiparticle, and
$P~(V)$ a pseudoscalar (vector) meson, is crucial for discovering excited baryons predicted by effective field theories but yet to be observed~\cite{baryons,excited}. Notably, experimental data on excited hyperon resonances, such as $\Lambda^{*}$ and $\Sigma^{*}$, are significantly sparser compared to those for baryons with zero strangeness~\cite{hyperon}. Therefore, experimental confirmation or exclusion of these predicted hyperons is essential for advancing hyperon spectroscopy.
   
In recent years, the BESIII collaboration has reported observations or evidence of several excited $\Lambda^{*}$ states via $\psi(3686)\to B\bar{B}P~(V)$ decays. For example, the $\Lambda(1670)$ is observed with a significance larger than 5.0$\sigma$ in the decay $\psi(3686)\to\Lambda\bar{\Lambda}\eta$~\cite{LLeta}. Evidence of an excited state
$\Lambda^{*}$ is found  in $\psi(3686)\to\Lambda\bar{\Lambda}\omega$ with a significance of 3.0$\sigma$~\cite{LLomega}. Additionally, a near-threshold enhancement in $\Lambda\bar{\Lambda} $ is observed for the first time in
$e^{+}e^{-}\to\Lambda\bar{\Lambda}\phi$~\cite{LLphi}. 
These findings motivate the search for excited $\Sigma^{*}$ states in the analogous three-body decay $\psi(3686)\to\Sigma^{0}\bar{\Sigma}^{0}\omega$.

In this paper, we report the first observation and branching fraction measurement of the decay
$\psi(3686)\to\Sigma^{0}\bar{\Sigma}^{0}\omega$, using $(27.12\pm 0.14)\times 10^{8}$ $\psi(3686)$
events~\cite{psip_num_21}. We also search for potential structures in the invariant mass distributions of  $\Sigma^{0}\omega$, $ \bar{\Sigma}^{0}\omega$ and $\Sigma^{0}\bar{\Sigma}^{0}$.

\section{BESIII Detector and Monte Carlo Simulation}\label{sec:detector}
\vspace{-0.3cm}

The BESIII detector~\cite{BES_design} records symmetric $e^+e^-$
collisions provided by the BEPCII storage
ring~\cite{Yu:IPAC2016-TUYA01} in the center-of-mass energy range from
1.84 to 4.95~GeV. The peak luminosity of $1.1\times10^{33}\,\text{cm}^{-2}\text{s}^{-1}$ is achieved at $\sqrt{s}=3.773\,\text{GeV}$. BESIII has collected large data samples in this
energy region~\cite{Ablikim:2019hff, EventFilter}. The
cylindrical core of the BESIII detector covers 93\% of the full solid
angle and consists of a helium-based multilayer drift chamber (MDC), a
time-of-flight system (TOF), and a CsI (Tl) electromagnetic
calorimeter (EMC), which are all enclosed in a superconducting
solenoidal magnet providing a 1.0~T magnetic field. Modules of the resistive plate muon
counter (MUC) are embedded in an octagonal flux-return yoke supporting
the superconducting solenoid. The charged-particle
momentum resolution at 1~GeV/$c$ is 0.5\%, and the specific ionization
energy loss d$E/$d$x$ resolution is 6\% for the electrons from Bhabha
scattering at 1~GeV. The EMC measures photon energy with a resolution
of 2.5\% (5\%) at 1~GeV in the barrel (end cap) region. The time
resolution of the TOF plastic scintillator barrel part is 68~ps, while
that of the end cap part was 110~ps. The end cap TOF system was
upgraded in 2015 using multigap resistive plate chamber technology,
providing a time resolution of 60~ps, which benefits $\sim83\%$ of
the data used in this analysis~\cite{tof_a,tof_b,tof_c}.

Monte Carlo (MC) simulated data samples produced with {\sc geant4}-based software~\cite{geant4}, which includes the geometric
description~\cite{detvis} of the BESIII detector and the detector response, are used to optimize the event selection criteria, determine
the detection efficiencies and study the background components. The simulation models the beam energy spread and initial-state radiation
in the $e^+e^-$ annihilations using the generator {\sc kkmc}~\cite{kkmc_a,kkmc_b}. The inclusive MC sample includes the
production of the $\psi(3686)$ resonance, the initial-state radiation production of the $J/\psi$, and the continuum processes incorporated
in {\sc kkmc}. All particle decays are modeled by {\sc evtgen}~\cite{evtgen_a, evtgen_b} using branching fractions either taken from the Particle Data Group (PDG)~\cite{pdg2022}, when available, or otherwise estimated with {\sc lundcharm}~\cite{lundcharm_a,lundcharm_b}. Final state radiation from charged final state particles is included using {\sc photos}~\cite{photos}. The signal MC sample of $\psi(3686)\to\Sigma^{0}\bar{\Sigma}^{0}\omega$, comprising $3.0\times 10^{6}$ events, is generated with an uniform phase space (PHSP) distribution. The decays $\Sigma^{0}\to \gamma \Lambda$($\Lambda \to p \pi^{-}$) and $\bar{\Sigma}^{0}\to\gamma\bar{\Lambda}$($\bar{\Lambda} \to \bar{p} \pi^{+}$) are also simulated using the PHSP model~\cite{Sigma0_pair}, and the decay of $\omega$ is modeled using {\sc omega\_dalitz} generator~\cite{evtgen_b}.
The data sample collected at the center-of-mass energy of $\sqrt{s}=3.650$~GeV, with an integrated luminosity of 401~pb$^{-1}$~\cite{psip_num_21}, is used to investigate the continuum background.

   \section{\label{Sec:Selection}Event Selection and Background Analysis}\vspace{-0.3cm}
   
To reconstruct the signal decay, we select the decay chains $\Sigma^{0} \to \gamma \Lambda$, $ \Lambda \to p
\pi^{-}$, $\bar{\Sigma}^{0} \to \gamma\bar{\Lambda}$,
$\bar{\Lambda}\to \bar{p} \pi^{+}$ and $\omega \to \pi^{+}\pi^{-}\pi^{0}$ with $\pi^{0} \to \gamma \gamma$. Each candidate event must contain at least three positive and three negative charged tracks and a minimum of four photons. Additionally, charged tracks detected in the MDC must satisfy $|\cos\theta|\leq 0.93$, where $\theta$ is the polar angle relative to the $z$-axis, the symmetry axis of the MDC.

Photon candidates are identified from showers in the EMC. The
deposited energy of each shower is required to exceed 25~MeV in the
barrel region ($|\cos\theta|<0.80$) or 50~MeV in the end
cap region ($0.86<|\cos\theta|<0.92$). To reduce electronic noise
and energy depositions unrelated to the event, the EMC cluster time, measured from the reconstructed event start time, must lie within $[0,700]$~ns.
To exclude showers that originate from
charged tracks,
the angle subtended by the EMC shower and the position of the closest charged track at the EMC
must be greater than 10~degrees as measured from the interaction point (IP).
    
Particle identification~(PID) for charged tracks combines measurements
of d$E$/d$x$ in the MDC and the flight time in the TOF to
calculate the probability $P(h)$ for each particle type
hypothesis, where $(h=p, K, \pi)$. Tracks are assigned as (anti-)proton candidates if they satisfy $P(p)>P(K)$ and $ P(p)>P(\pi)$. Each candidate event must include at least one proton and one anti-proton. To account for anti-proton interactions with detector materials, the opening angle between the anti-proton trajectory and any photon shower must exceed 20~degrees.

\begin{figure*}[tb]
    \centering
    \mbox{
    \begin{overpic}[width=0.4\textwidth,clip=true]{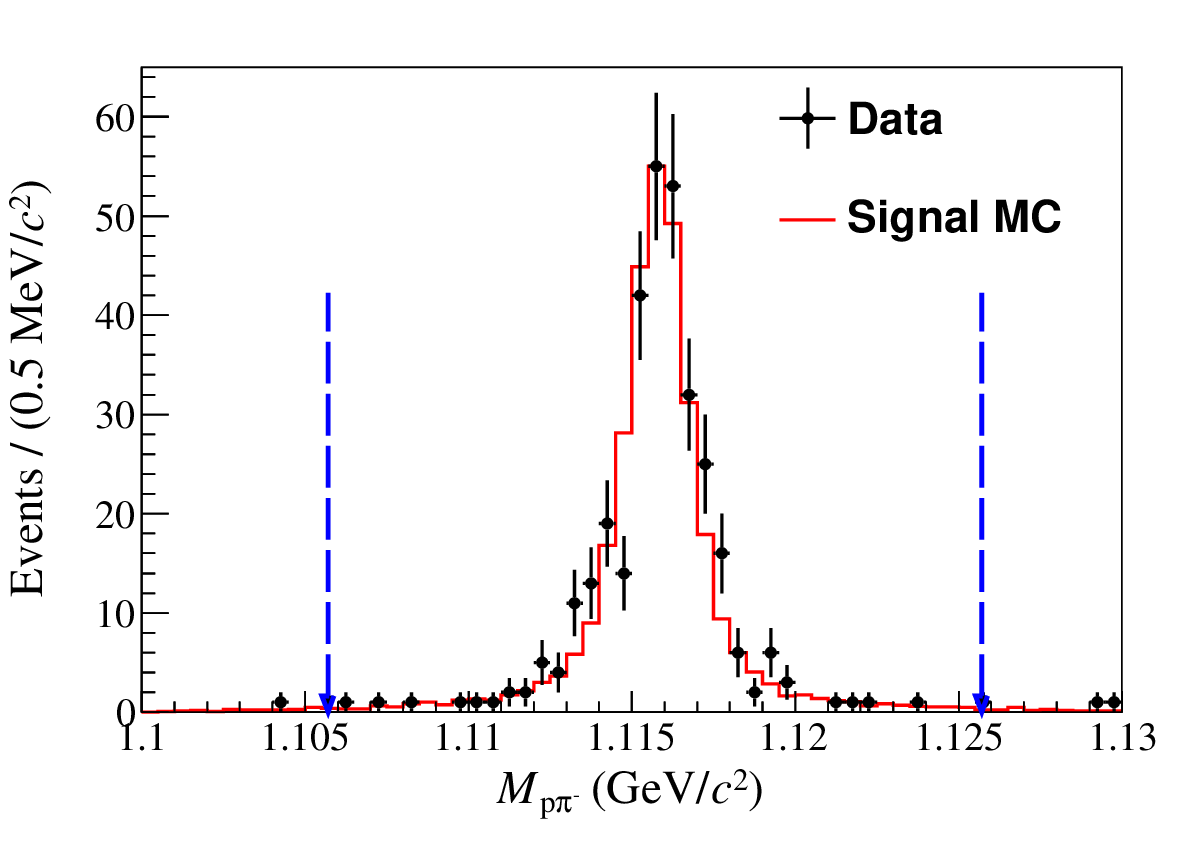}
    \end{overpic}
    \begin{overpic}[width=0.4\textwidth,clip=true]{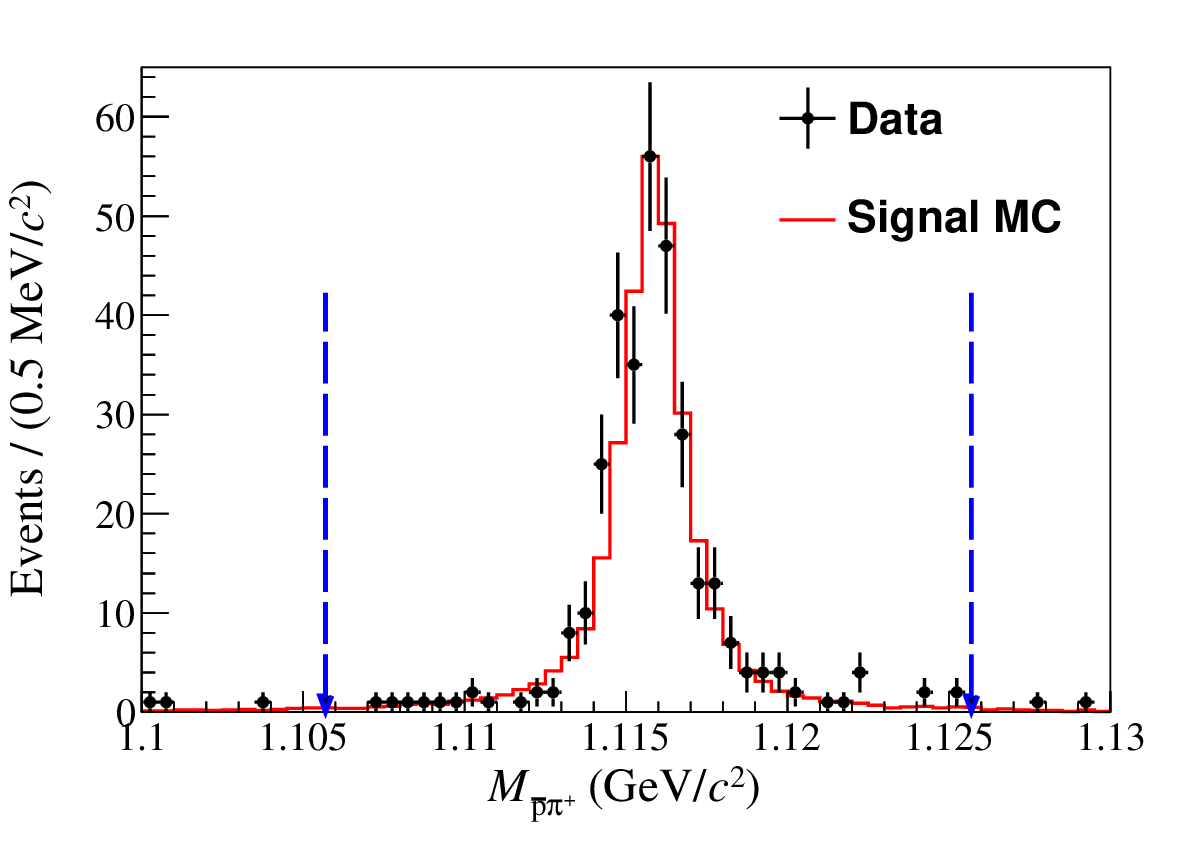}
    \end{overpic}
    }
    \caption{Distributions of (left) $M_{p\pi^{-}}$ and (right) $M_{\bar{p}\pi^{+}}$. The dots with error bars are data, the red solid lines represent the normalized signal MC sample, and the blue dashed lines indicate the required mass ranges.}
   \label{mlamb}
   \end{figure*}

The $\Lambda$$(\bar{\Lambda})$ candidates are reconstructed by combining identified (anti-)protons with oppositely charged tracks and performing a secondary vertex fit. If multiple combinations satisfy the fit, the one is selected with minimizing $(M_{p\pi^{-}} - m_{\Lambda})^{2}+(M_{\bar{p}\pi^{+}} - m_{\bar{\Lambda}})^{2}$, where $M_{p\pi^{-}}$($M_{\bar{p}\pi^{+}}$) is the invariant mass of the $p\pi^{-}$($\bar{p}\pi^{+}$)
pair, and $m_{\Lambda}$($m_{\bar{\Lambda}}$) is the nominal mass~\cite{pdg2022} of $\Lambda$($\bar{\Lambda}$). 
The decay length of $\Lambda$($\bar{\Lambda}$), defined as the distance from the $\Lambda$($\bar{\Lambda}$) decay vertex to the IP, must be greater than zero. Reconstructed $\Lambda$ and $\bar{\Lambda}$ candidates must satisfy $\left|M_{p\pi^{-}(\bar{p}\pi^{+})}-m_{\Lambda(\bar{\Lambda})}\right|<0.01$~GeV/$c^2$~(five standard deviations), as illustrated in Fig.~\ref{mlamb}.

For the remaining charged tracks not involved in reconstructing $\Lambda$ and $\bar{\Lambda}$, we assume they are pions originating from the $\omega$ decay. These tracks must satisfy $\left|V_{z}\right|<10$~cm and
$\left|V_{xy}\right|<1$~cm, where $\left|V_{z}\right|$ is the
distance of closest approach to the IP along the $z$ direction and $V_{xy}$ is the distance in the transverse $xy$~plane. A vertex fit is performed on each $\pi^{+}\pi^{-}$ pairs to ensure a common decay vertex.

In order to improve the invariant mass resolution and reduce the backgrounds, a five-constraint (5C) kinematic fit is applied to each $\psi(3686)\to \Lambda\bar{\Lambda}\pi^{+}\pi^{-}4\gamma$ candidate, enforcing energy-momentum conservation and constraining photon pairs to the nominal $\pi^{0}$ mass.
The combination with the smallest $\chi^{2}_{5\mathrm{C}}$ is selected, requiring $\chi^{2}_{5\mathrm{C}}<60$. This criterion is optimized using a figure-of-merit defined as $\frac{S}{\sqrt{S+B}}$, where $S$ is the normalized number of signal events in the signal MC sample and $B$ is the number of background events in the inclusive MC sample. 
To suppress backgrounds with different photon multiplicities than the signal decay, a four-constraint (4C) kinematic fit is applied to $\Lambda\bar{\Lambda}\pi^{+}\pi^{-}4\gamma$, $\Lambda\bar{\Lambda}\pi^{+}\pi^{-}3\gamma$ and $\Lambda\bar{\Lambda}\pi^{+}\pi^{-}5\gamma$ combinations (if more than four photons are present in the event), requiring $\chi^{2}_{4\mathrm{C}}(\Lambda\bar{\Lambda}\pi^{+}\pi^{-}4\gamma)<\chi^{2}_{4\mathrm{C}}(\Lambda\bar{\Lambda}\pi^{+}\pi^{-}3\gamma)$ and $\chi^{2}_{4\mathrm{C}}(\Lambda\bar{\Lambda}\pi^{+}\pi^{-}4\gamma)<\chi^{2}_{4\mathrm{C}}(\Lambda\bar{\Lambda}\pi^{+}\pi^{-}5\gamma)$.

In the signal decay, the $\Sigma^{0}$ and $\bar{\Sigma}^{0}$ states always appear in pairs and the relevant photons are expected to yield $M_{\gamma^{i}\Lambda}$ close to $M_{\gamma^{j}\bar{\Lambda}}$, where $M_{\gamma^{i}\Lambda}$ and $M_{\gamma^{j}\bar{\Lambda}}$ are
the invariant masses of $\gamma^{i}\Lambda$ and $\gamma^{j}\bar{\Lambda}$, respectively. Here, $\gamma^{i}$ and $\gamma^{j}$ are different photons not involved in reconstructing $\pi^{0}$. We select the photon candidate by minimizing the variable $\Delta=(M_{\gamma^{i}\Lambda}-M_{\gamma^{j}{\Lambda}})^{2}$.

To veto backgrounds from the decay $\psi(3686)\to\pi^{+}\pi^{-}(\pi^{0}\pi^{0})J/\psi$, we require the recoil mass of $\pi^{+}\pi^{-}(\pi^{0}\gamma^{i}\gamma^{j})$ to be outside of the $J/\psi$ mass window. The mass window is set at 
[3.089, 3.105]~GeV/$c^{2}$ for $\psi(3686)\to\pi^+\pi^-J/\psi$ and adjusted to [3.090, 3.145]~GeV/$c^{2}$ for $\psi(3686)\to\pi^0\pi^0J/\psi$.
   
Background candidates are studied using the $\psi(3686)$ inclusive
MC sample with {\tt TopoAna} package~\cite{zhouxy_topoAna}. Within the fit range of [0.65,0.95]~GeV/$c^{2}$, no peaking background is observed in the $M_{\pi^{+}\pi^{-}\pi^{0}}$ spectrum. The remaining dominant background events are $\psi(3686)\to\pi^{+}\pi^{-}J/\psi,J/\psi \to \pi^{0}\Sigma^{0}\bar{\Sigma}^{0}$. Continuum background candidates, estimated using off-resonance data at $\sqrt{s}=3.650$~GeV, are negligible.

   \section{\label{Sec:BR_determined}Branching Fraction Measurement}\vspace{-0.3cm}
   
\begin{figure}[tb]
    \centering
    \mbox{
    \begin{overpic}[width=0.4\textwidth,clip=true]{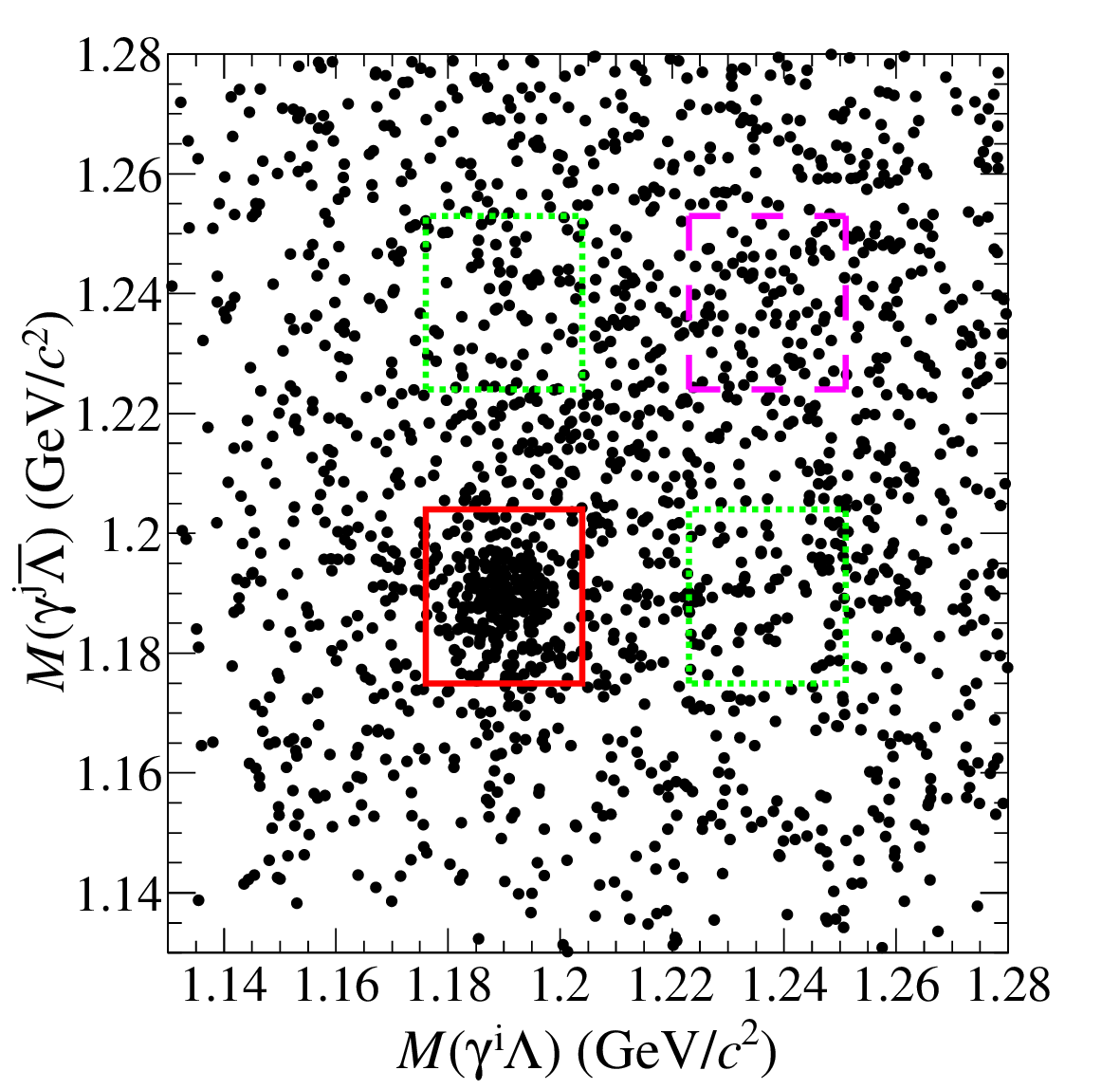}
    \end{overpic}
    }
    \caption{Distribution of $M_{\gamma^{i}\Lambda}$ versus $M_{\gamma^{j}\bar{\Lambda}}$ in the data. Events are within the $\omega$ mass window. The red solid box indicates the signal region, the green dashed boxes denote the 1-D sideband regions, and the magenta long dashed box represents the $\Sigma^{0}-\bar{\Sigma}^{0}$ sideband region.}
   \label{SS_scatter}
   \end{figure}

After applying all selection criteria, we fit the spectra of $M_{\gamma^{i}\Lambda}$, $M_{\gamma^{j}\bar{\Lambda}}$ and $M_{\pi^{+}\pi^{-}\pi^{0}}$ to determine the signal and sideband regions, as detailed in Table~\ref{mass_windows}.
 \begin{table}[tb]
        \centering        
        \caption{Signal and sideband regions of $\Sigma^{0}$, $\bar{\Sigma}^{0}$ and $\omega$.}
        \label{mass_windows}
        \begin{tabular}{ccc}
        \hline
        \hline
               &Signal region (GeV/$c^2$)&Sideband region (GeV/$c^2$) \\
             \hline
           $M_{\gamma^{i}\Lambda}$  & $[1.176, 1.204]$ & $[1.223, 1.251]$\\
            $M_{\gamma^{j}\bar{\Lambda}}$  & $[1.175, 1.204]$ &$[1.224, 1.253]$\\
             $M_{\pi^{+}\pi^{-}\pi^{0}}$& $[0.758, 0.822]$&$[0.630, 0.694]$, $[0.886, 0.950]$ \\

        \hline
        \hline
        \end{tabular}
    \end{table}
The signal events of $\Sigma^{0}$ and $\bar{\Sigma}^{0}$ are distinctly visible in the scatter plot of $M_{\gamma^{i}\Lambda}$ versus $M_{\gamma^{j}\bar{\Lambda}}$, as illustrated in Fig.~\ref{SS_scatter}.
   
The signal yield of the decay $\psi(3686)\to\Sigma^{0}\bar{\Sigma}^{0}\omega$ is extracted using an extended unbinned maximum likelihood fit~\cite{extend_likelihood} to the $M_{\pi^{+}\pi^{-}\pi^{0}}$ spectrum, considering only events within the 2-D $\Sigma^{0}$$\bar{\Sigma}^{0}$ signal region. The $\omega$ signal is described with the signal MC simulated shape~\cite{keyspdf} and the smooth background contribution is modeled with a second-order Chebyshev polynomial function.
     \begin{figure}[b]
    \centering
    \mbox{
    \begin{overpic}[width=0.45\textwidth,clip=true]{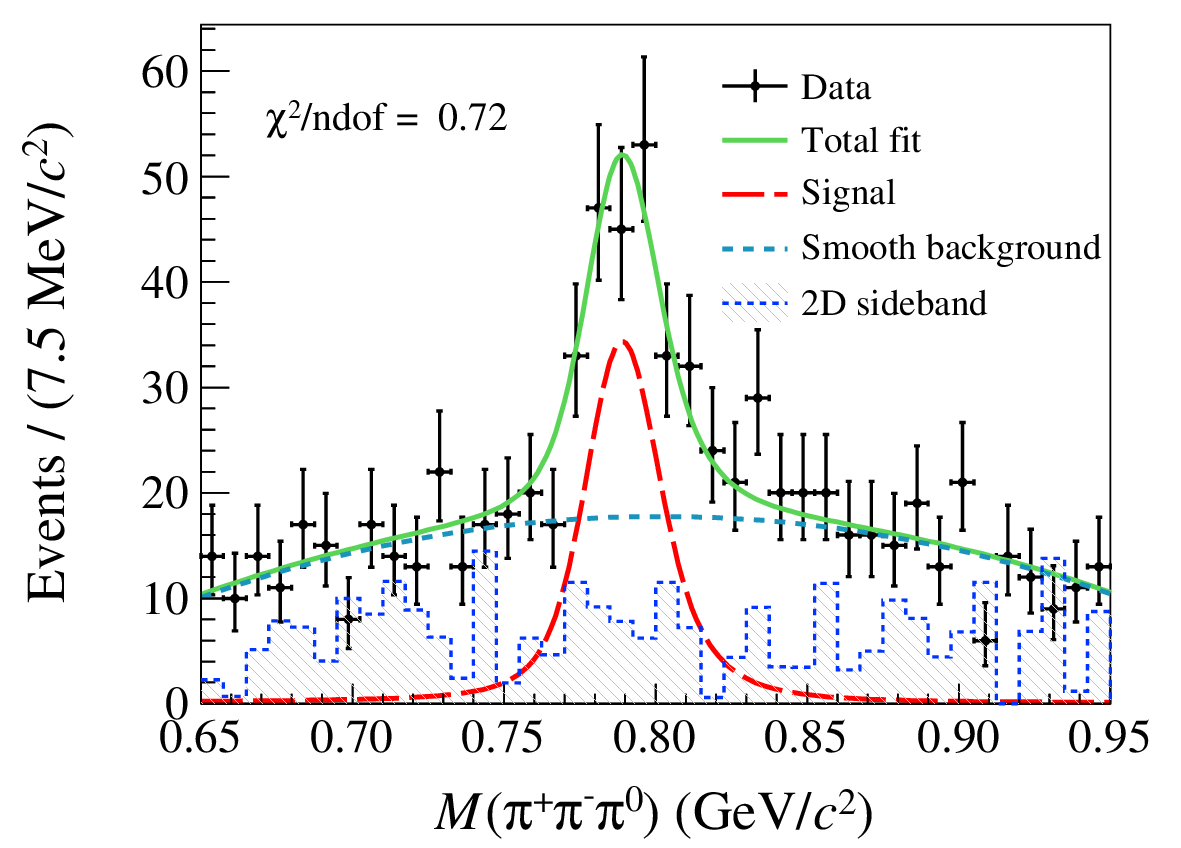}
    \end{overpic}
    }
    \caption{Fit to the $M_{\pi^{+}\pi^{-}\pi^{0}}$ distribution. The dots with error bars are data, the grass green solid line denotes the fit result, the red long dashed line represents the signal component, the blue dashed line is smooth background, and the light blue shadowed area represents the smooth 2-D sideband of $\Sigma^{0}$ and $\bar{\Sigma}^{0}$.}
   \label{fit_result}
   \end{figure}
   
The fit result is shown in Fig.~\ref{fit_result}. The events from the 2-D sideband are overlayed to confirm the absence of any peaking background contribution. From this fit, the number of signal events is determined to be 171.7$\pm$22.3. The statistical significance, estimated from the difference in the log-likelihood values with and without the signal component, is 8.9$\sigma$. The branching fraction is calculated as
   \begin{equation}
   \Br(\psi(3686)\to\Sigma^{0}\bar{\Sigma}^{0}\omega) =
    \frac{{N}_{\mathrm{sig}}^{\mathrm{obs}}}{N_{\psi \left( {3686} \right)} 
   \cdot \varepsilon_{\mathrm {sig}} \cdot \Br_{\omega}},
   \end{equation}
where ${N_{\mathrm{sig}}^{\mathrm{obs}}}$ is the signal yield, $\Br_{\omega}$ is the branching fraction of $\omega \to \pi^{+}\pi^{-}\pi^{0}$~\cite{pdg2022},
$\varepsilon_{\mathrm{sig}}$ represents the signal efficiency determined by MC simulation, and $N_{\psi \left(3686\right)}=(27.12\pm0.14)\times10^{8}$~\cite{psip_num_21} is the total number of $\psi(3686)$ events in the data. The branching fraction of
$\Sigma^{0}\to\gamma\Lambda$ is 100\% according to the PDG, and therefore not included in Eq.~(1). The branching fractions of $\Lambda\to p\pi^{-}$, $\bar{\Lambda}\to\bar{p}\pi^{+}$ and $\pi^{0}\to\gamma\gamma$ are incorporated into the signal efficiency. Table~\ref{br_result} includes the numerical results of the branching fraction measurement.
    \begin{table}[tb]
        \centering        
        \caption{Signal yield, statistical significance, efficiency and measured branching fraction. The uncertainties are statistical only.}
   \label{br_result}
\begin{tabular}{lcccc}
\hline
\hline
&Signal yield& Significance & $\varepsilon(\%)$ & $\Br~(\times10^{-5})$ \\
\hline
    &  171.7$\pm$22.3 &8.9$\sigma$        & 0.57   & 1.24$\pm$0.16    \\

\hline
\hline
\end{tabular}
\end{table}

   \section{\label{Sec:inter_states}Study of Intermediate States}\vspace{-0.3cm}

\begin{figure*}[tb]
   \centering
   \begin{minipage}[t]{0.32\linewidth}
   \includegraphics[width=1\textwidth]{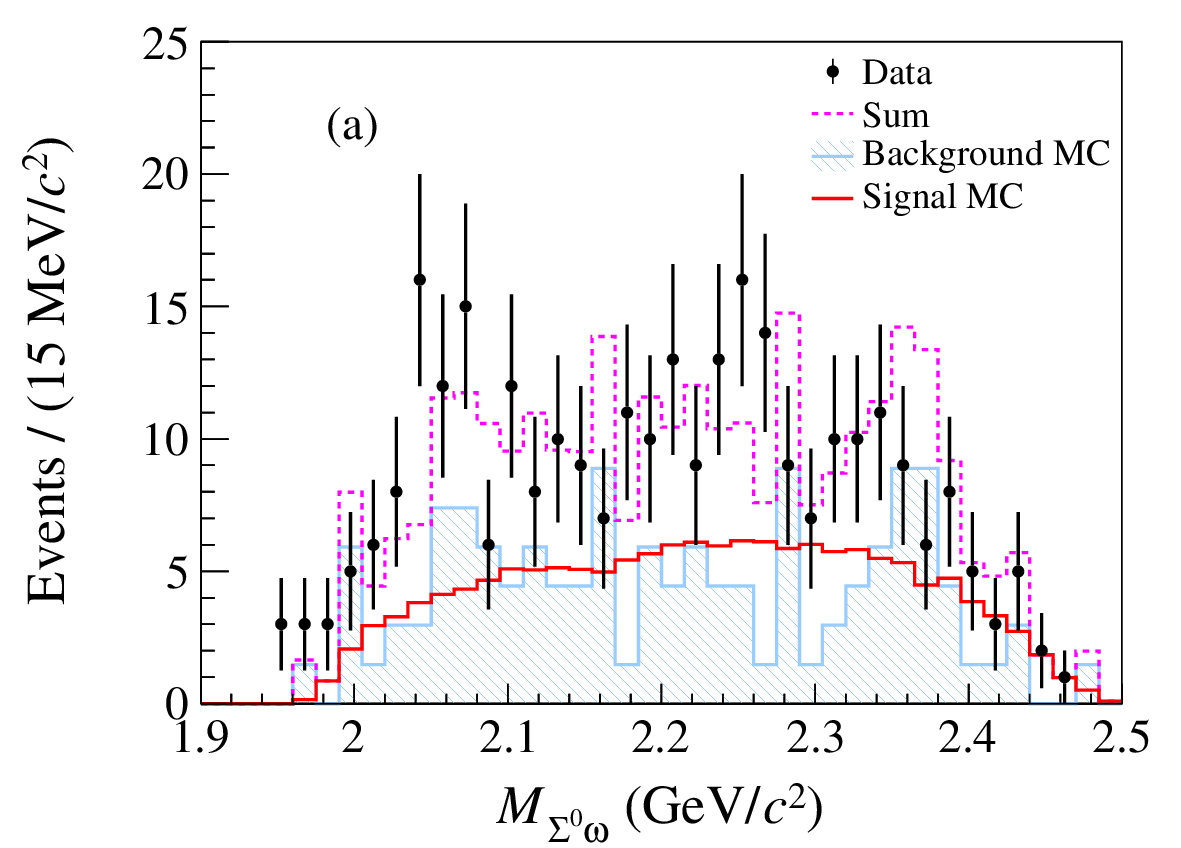}
  \end{minipage}
  \begin{minipage}[t]{0.32\linewidth}
  \includegraphics[width=1\textwidth]{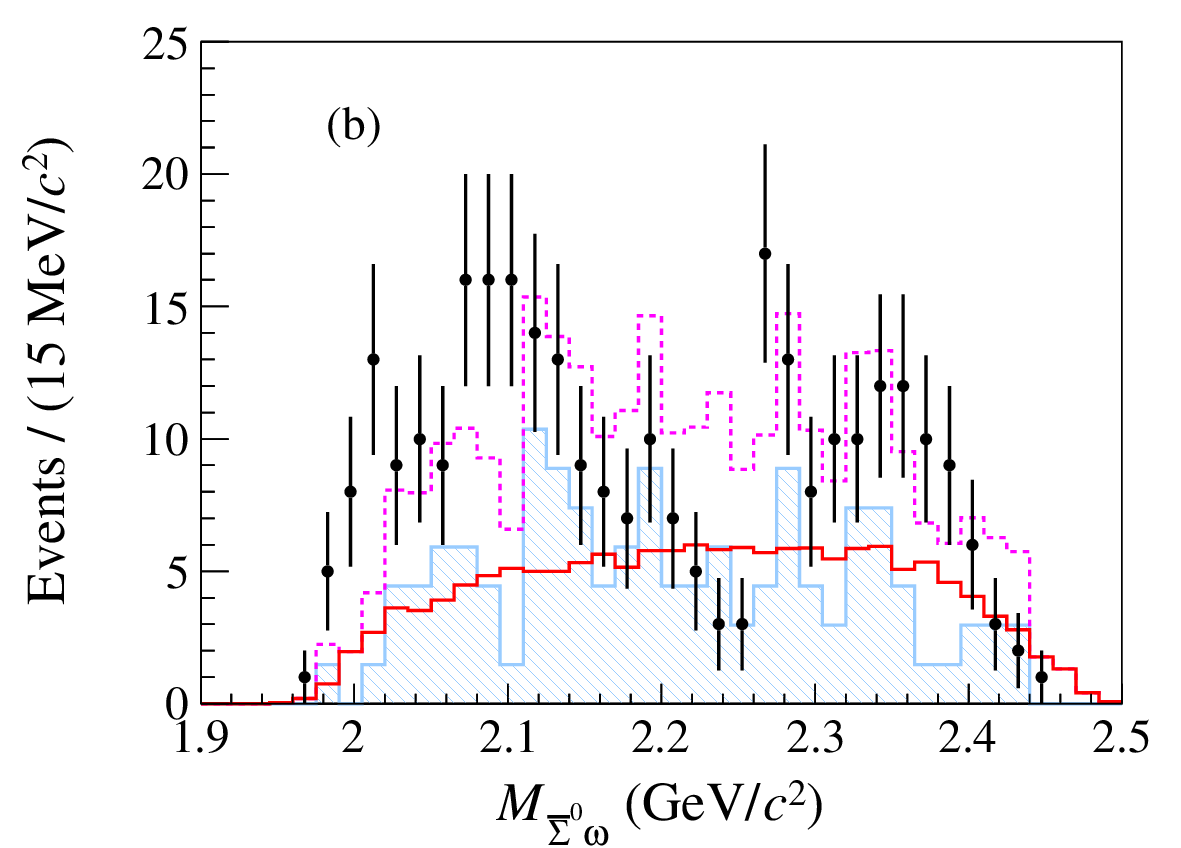}
  \end{minipage}
  \begin{minipage}[t]{0.32\linewidth}
  \includegraphics[width=1\textwidth]{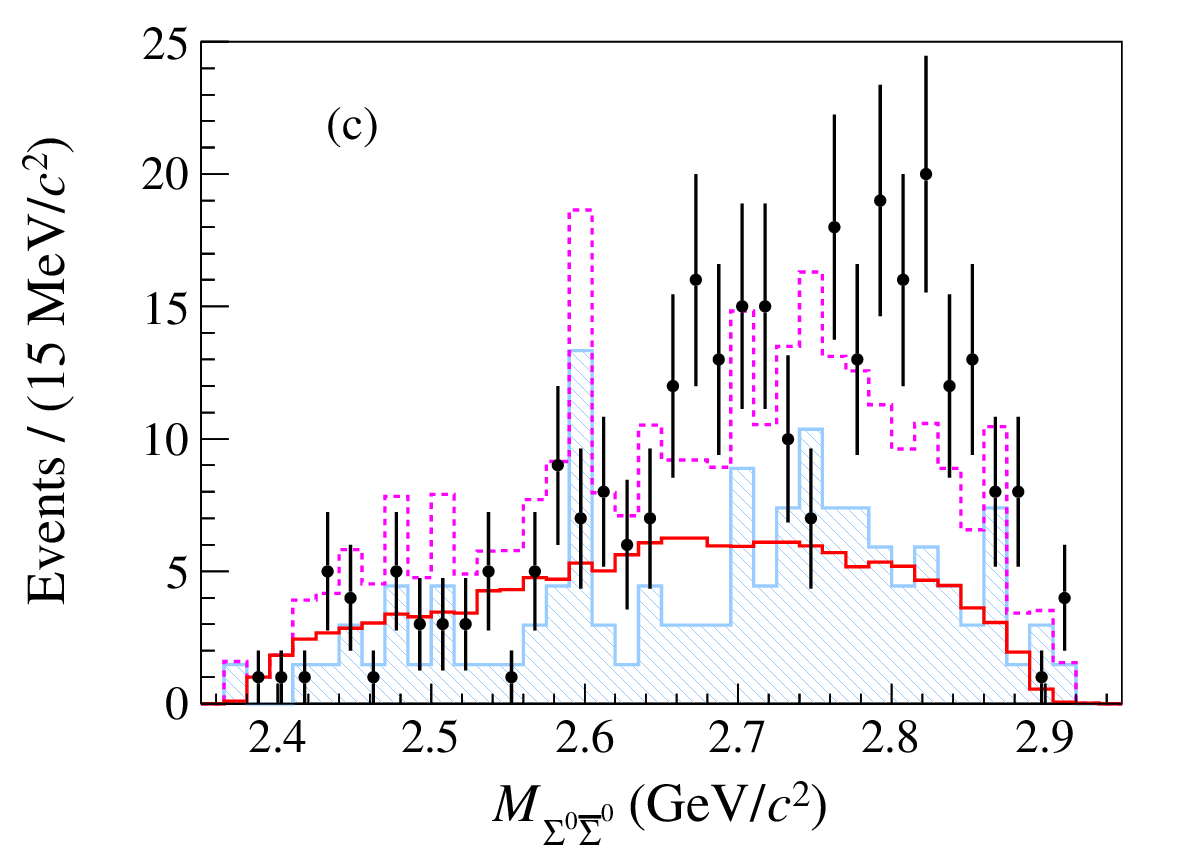}
  \end{minipage}
  
  \begin{minipage}[t]{0.32\linewidth}
  \includegraphics[width=1\textwidth]{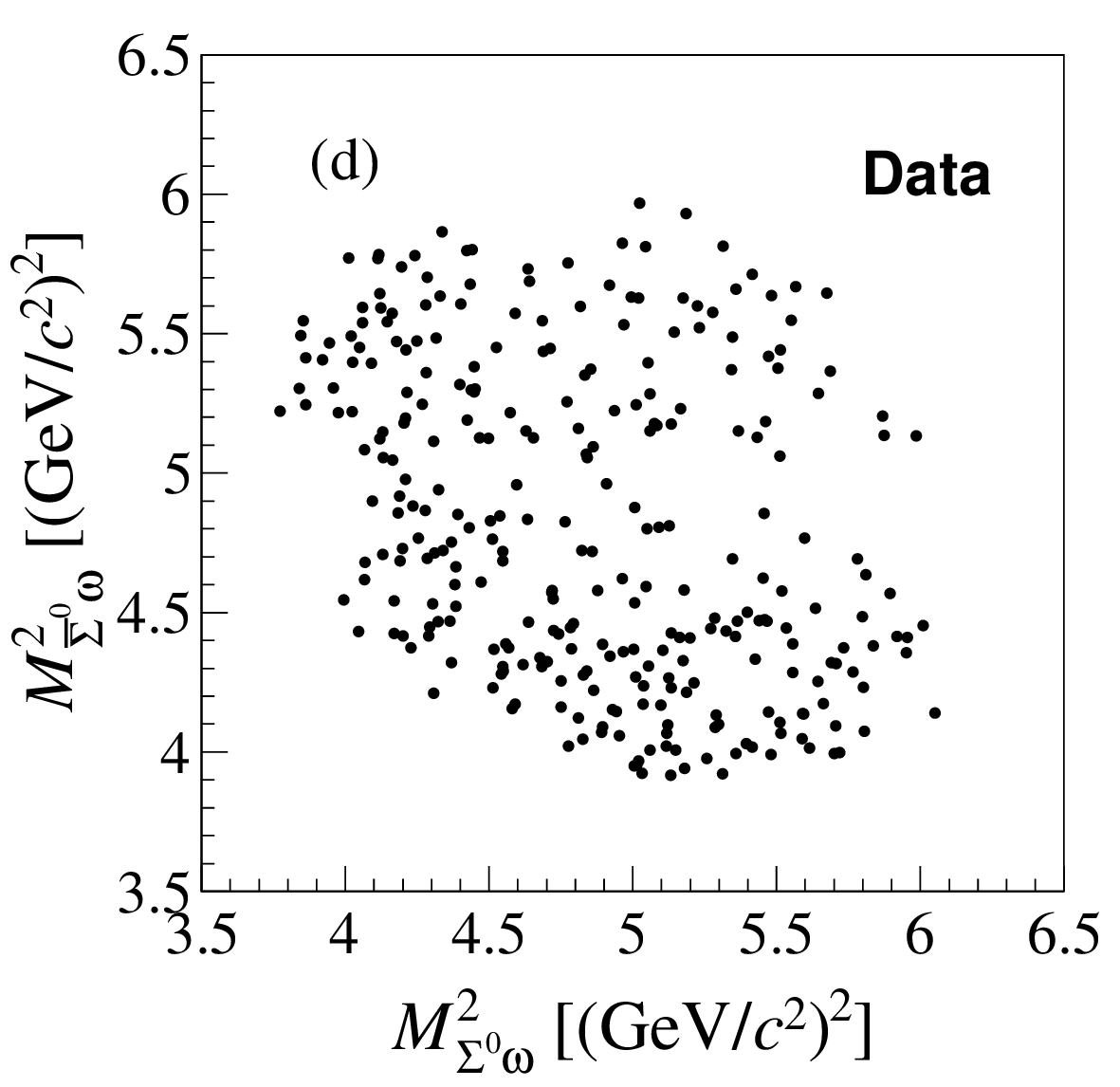}
  \end{minipage}
  \begin{minipage}[t]{0.32\linewidth}
  \includegraphics[width=1\textwidth]{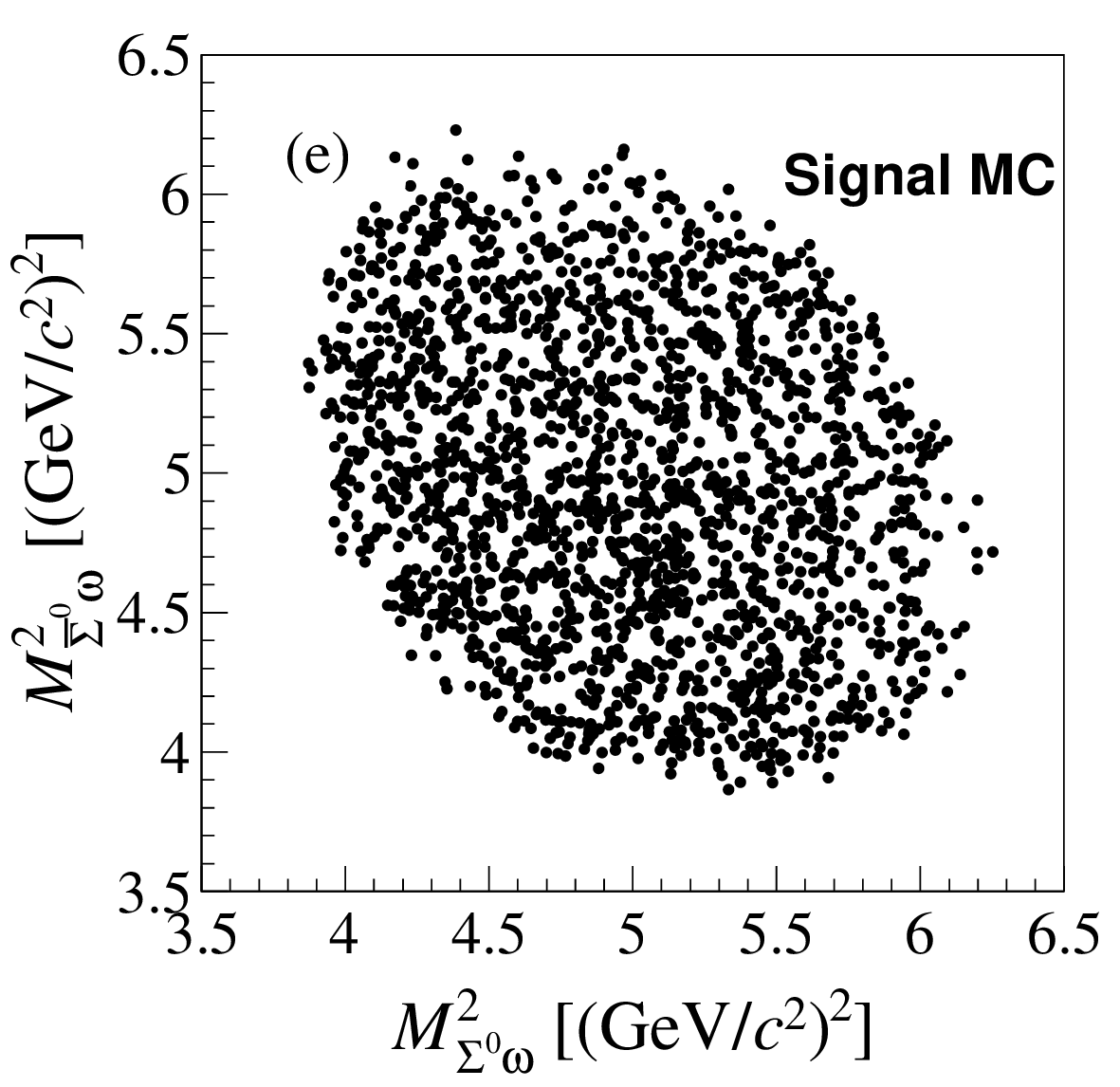}
  \end{minipage}

   \caption{(top) Distributions of (a) $M_{\Sigma^{0}\omega}$,
     (b) $M_{\bar{\Sigma}^{0}\omega}$ and (c) $M_{\Sigma^{0}\bar{\Sigma}^{0}}$ in the data and MC samples. The black dots with error bars represent the data, the red solid lines indicate the PHSP signal MC simulated shapes, the shaded blue areas depict the simulated background events derived from the inclusive MC sample and the magenta dashed lines represent the sum of signal and background MC shapes. The numbers of signal and background events are normalized to their respective values in the data. (bottom) Dalitz plot of ${M^{2}_{\Sigma^{0}\omega}}$ versus
     ${M^{2}_{\bar{\Sigma}^{0}\omega}}$ in the (d) data and (e) signal MC sample.}
   \label{inter_plots}
   \end{figure*}

The distributions of $M_{\Sigma^{0}\omega}$, $M_{\bar{\Sigma}^0\omega}$ and
$M_{\Sigma^{0}\bar{\Sigma}^0}$, along with the Dalitz plot of
${M^{2}_{\Sigma^{0}\omega}}$ versus ${M^{2}_{\bar{\Sigma}^{0}\omega}}$ presented in Fig.~\ref{inter_plots} are used to explore the potential intermediate states.
Differences between the data and the MC simulation are observed in the $M_{\Sigma^{0}\omega}$ and $M_{\bar{\Sigma}^{0}\omega}$ distributions, as well as in the uneven distribution of the Dalitz plot. 

\begin{figure}[tb]
    \centering
    \mbox{
    \begin{overpic}[width=0.45\textwidth,clip=true]{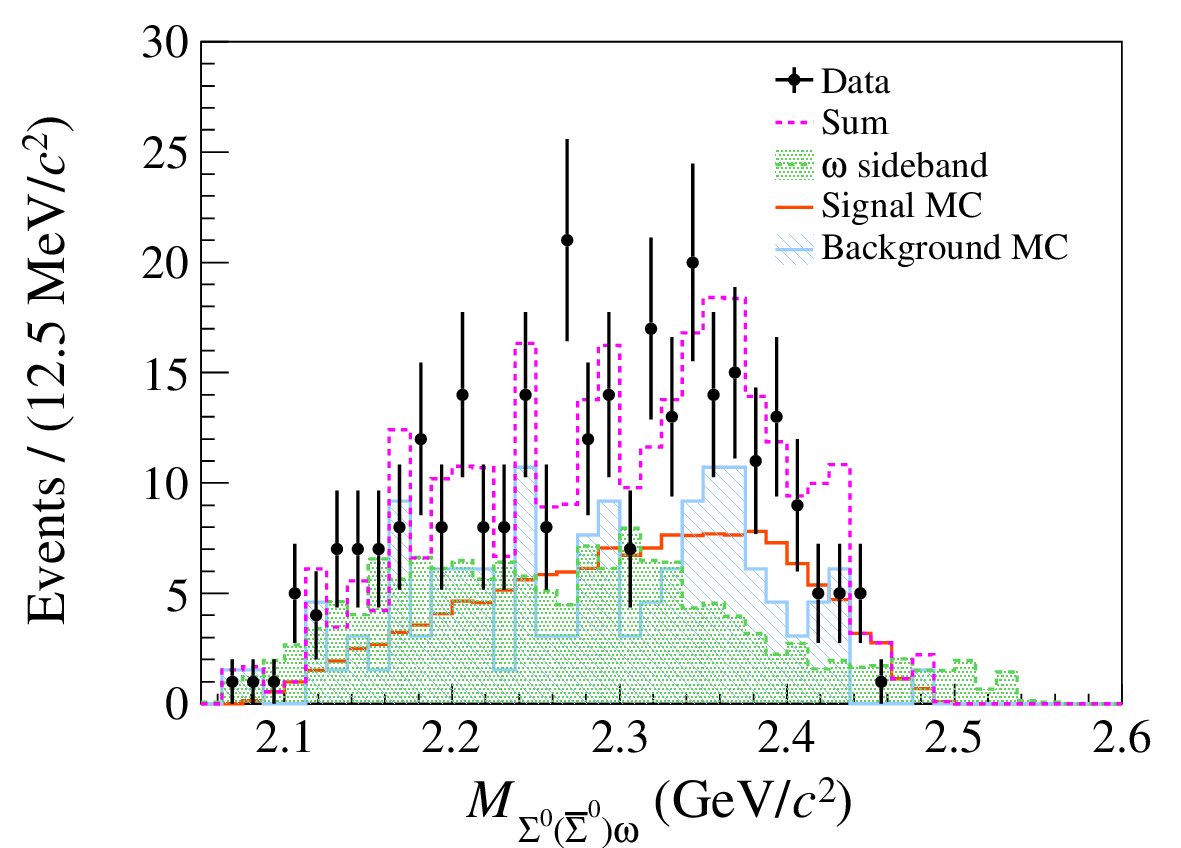}
    \end{overpic}
    }
    \caption{Distributions of $M_{\Sigma^{0}(\bar{\Sigma}^0)\omega}$ obtained by retaining the larger value between $M_{\Sigma^{0}\omega}$ and $M_{\bar{\Sigma}^{0}\omega}$ in the data and MC samples. The dots with error bars are the data, the green shaded histogram indicates the $\omega$ sideband, the light blue shaded histogram represents the simulated backgrounds derived from the inclusive MC sample, the red solid line depicts the PHSP signal MC shape and the magenta dashed line represents the sum of signal and background MC shapes. The numbers of events are normalized to the data determined by the fit to the $M_{\pi^{+}\pi^{-}\pi^{0}}$ spectrum.}
   \label{overlap}
   \end{figure}

\begin{figure}[tb]
    \centering
    \mbox{
    \begin{overpic}[width=0.45\textwidth,clip=true]{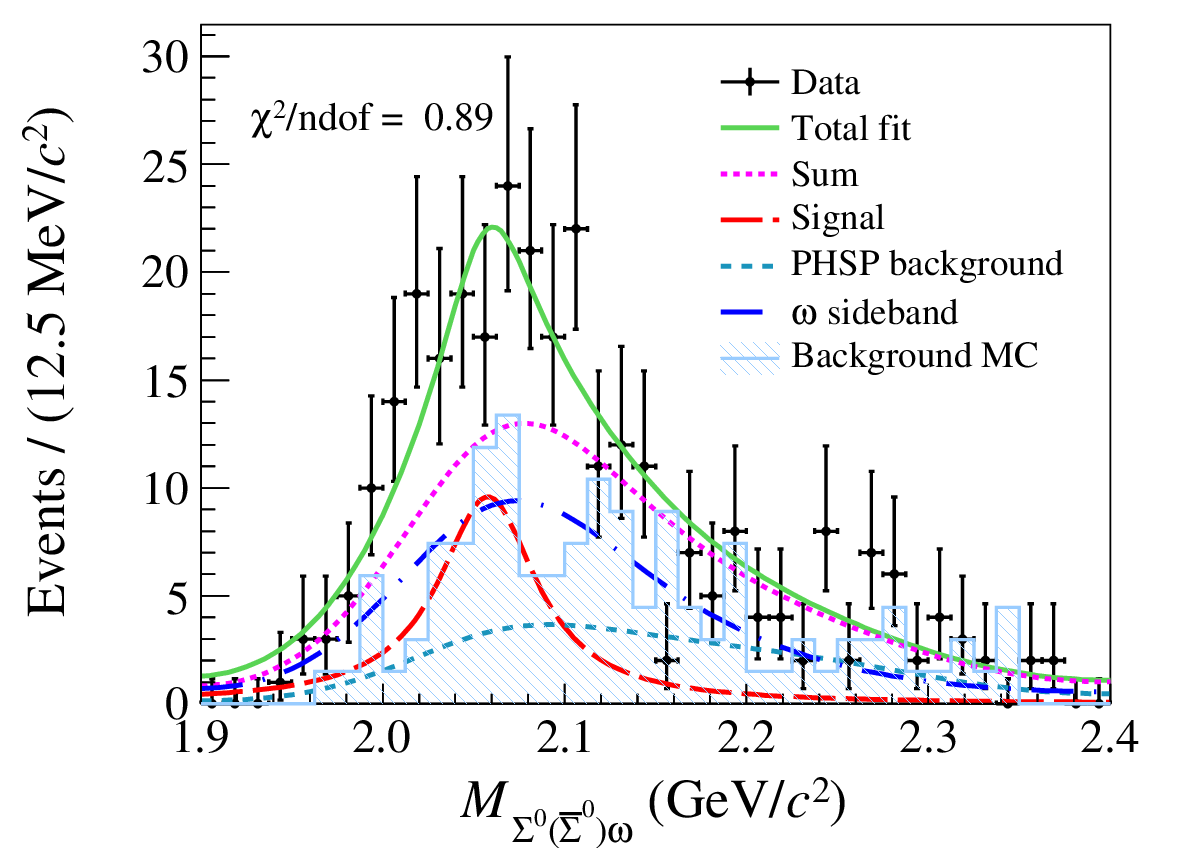}
    \end{overpic}
    }
    \caption{Fit to the $M_{\Sigma^{0}(\bar{\Sigma}^0)\omega}$ spectrum, retaining the smaller mass value. The grass green solid line is the fit result, the red long dashed line represents the signal function, the dark blue long dash-dotted line depicts the $\omega$ sideband, the light blue dashed line indicates the PHSP MC shape, the magenta short dashed line represents the sum of the sideband and PHSP MC shape, and the light blue shaded histogram represents the simulated backgrounds.}
   \label{fit_inter_states}
   \end{figure}

To extract potential structure in the $M_{\Sigma^{0}\omega}$ and $M_{\bar{\Sigma}^{0}\omega}$ distributions, it is crucial to account for reflections from the charge-conjugate channels. Particularly, we compare $M_{\Sigma^{0}\omega}$ and $M_{\bar{\Sigma}^0\omega}$ in each event and retain only one of them, denoted as $M_{\Sigma^{0}(\bar{\Sigma}^0)\omega}$. First, we examine the distribution for the larger value of $M_{\Sigma^{0}\omega}$ or $M_{\bar{\Sigma}^0\omega}$ (Fig.~\ref{overlap}) finding no evidence of a structure. Later, the distribution of $M_{\Sigma^{0}\omega}$ or $M_{\bar{\Sigma}^{0}\omega}$ corresponding to the smaller value (Fig.~\ref{fit_inter_states}) is retained for further analysis.

In Fig.~\ref{fit_inter_states}, a structure near 2.06~GeV/$c^{2}$ is evident. To analyze this feature, an unbinned extended maximum likelihood fit is applied to the $M_{\Sigma^{0}(\bar{\Sigma}^0)\omega}$ spectrum. 
In the fit, the PHSP shape is derived from the exclusive MC simulation of $\psi(3686)\rightarrow\Sigma^{0}\bar{\Sigma}^{0}\omega$. The signal function is modeled by a relativistic Breit-Wigner function, defined as
\begin{equation}
    \left|BW(M)\right|^{2}=\frac{k}{{(M-m_{0})}^{2}+m^{2}_{0}\Gamma^{2}},
\end{equation}
where $k$ is a normalization factor, $M$ corresponds to $M_{\Sigma^{0}(\bar{\Sigma}^0)\omega}$, $m_{0}$ and $\Gamma$ are the mass and width of the resonance, respectively. For simplicity, the relatively smooth efficiency curve of the signal process is not incorporated into the signal function. The number of non-$\omega$ events is fixed according to the previous fit result. Two options are considered to describe the background contribution: a sum of background candidates in the inclusive MC sample, which yields a significance of $5.1\sigma$; or the $\omega$ sideband from the data, which yields a significance of $3.1\sigma$. To be conservative, we adopt the latter as the nominal one. The simulated background events are not included in the fit. It is presented in Fig.~\ref{fit_inter_states} for comparison. In order to estimate the significance of the intermediate state, we consider uncertainties arising from the background shape, the signal function, and the number of non-$\omega$ events. To take into account the uncertainties, the alternative fits are performed. The background events obtained from the $\omega$ sideband are changed by increasing the sideband range to 1.5 times its original width. The signal function is smeared with a free Gaussian function. The number of non-$\omega$ events, fixed in the fit, is varied by one standard deviation. As a conservative estimation, the smallest significance among these fits is retained. By taking into account the systematic uncertainties and the look-elsewhere effect~\cite{LEE}, the global significance of this resonance is 2.5$\sigma$.

According to the fit, the mass and width of this structure are (2058 $\pm$ 15)~MeV/$c^{2}$ and (65 $\pm$ 18)~MeV, respectively. Since its significance is less than 3$\sigma$, we refrain from reporting an upper limit for the branching fraction of $\Br({\psi(3686)
\rightarrow \Sigma^{0}X + \mathrm{c.c.}\, \to \Sigma^{0}\bar{\Sigma}^{0} \omega)}$, where $X$ denotes the potential intermediate state.

The observed deviation in the $M_{\Sigma^{0}(\bar{\Sigma}^{0})\omega}$ spectrum suggests the possible presence of an intermediate state. However, the current data statistics is insufficient to confirm its existence or to determine its quantum numbers. Detailed studies with larger datasets and improved model are necessary to verify this resonance.

   \section{Systematic Uncertainties}\label{sec:sysU}\vspace{-0.3cm}
   
   The sources of systematic uncertainty in the branching fraction measurement are described as below and summarized in Table~\ref{uncertainty}. Assuming all uncertainties arise from independent sources, the total systematic uncertainty is determined to be 9.2\%, by summing the individual contributions in quadrature.

The uncertainty associated with pion tracking is assigned as 1.0\% per pion, according to the studies of the control samples $J/\psi \to K^{*}\bar{K}$ and $J/\psi\to p\bar{p}\pi^{+}\pi^{-}$~\cite{pion_tracking}. The uncertainty is only assigned to the charged pions originating from the $\omega$.

The uncertainty from photon reconstruction is assigned as 0.5\% per photon, based on the study of the control sample $e^{+}e^{-}\to\gamma\mu^{+}\mu^{-}$~\cite{photon_rec}.
 
 The uncertainty related to $\pi^{0}$ reconstruction is momentum-dependent, following the relation $(0.06-2.41\times\frac{p}{\mathrm{GeV}/c})$\% for a single $\pi^{0}$, where $p$ is the $\pi^{0}$ momentum. This relation is obtained with the control samples $\psi(3686)\to \pi^{0}\pi^{0}J/\psi$ and $e^{+}e^{-}\to \omega \pi^{0}$ at $\sqrt{s}=3.773$~GeV~\cite{SSphi}. In the signal decay, the momentum of the $\pi^{0}$ is distributed around 0.3~GeV/$c$. The momentum-weighted uncertainty, 1.0\%, is assigned as the systematic uncertainty.

The differences in $\Lambda$ and $\bar{\Lambda}$ reconstruction efficiency, including the tracking and PID of $p\pi$, as well as their decay length and mass windows, are studied using the control sample $J/\psi\rightarrow p K^{-} \bar{\Lambda} + \mathrm{c.c.}$~\cite{rec_lam(lamb)}. After correcting the signal efficiencies based on the data-MC difference, the residual uncertainties are assigned as the corresponding systematic uncertainties, each being 0.9\% for both $\Lambda$ and $\bar{\Lambda}$. Assuming the uncertainties of $\Lambda$ and $\bar{\Lambda}$ reconstruction to be fully correlated, the overall systematic uncertainty due to $\Lambda$ and $\bar{\Lambda}$ reconstruction is determined to be 1.8\%.

The uncertainty associated with the 5C kinematic fit is estimated by comparing the efficiency with and without helix parameter corrections, following the method described in Ref.~\cite{5C_method}. The difference in the signal efficiency, 1.1\%, is assigned as the systematic uncertainty.

   The uncertainty of the correlated 2-D $\Sigma^{0}$$\bar{\Sigma}^{0}$ mass window is estimated using the control sample of $J/\psi\to \Sigma^{0}\bar{\Sigma}^{0}$. The difference in the acceptance efficiency between the data and MC simulation, 0.1\%, is account for the systematic uncertainty. The uncertainties associated with vetoing $\psi(3686)\to\pi^{+}\pi^{-}(\pi^{0}\pi^{0})J/\psi$ events are estimated using the Barlow test~\cite{Barlow_test}. We modify the nominal $J/\psi$ mass windows by reducing or expanding them five times with increments of $\pm2$~MeV/$c^{2}$, then re-measure the branching fraction for each adjusted mass window. The test variable is defined as $\zeta=\frac{\left|V_{\text{nominal}}-V_{\text{test}}\right|}{\sqrt{\left|\sigma^2_{\mathrm{V} \text{nominal}}-\sigma^2_{\mathrm{V} \text{test}}\right|}}$, where $V_{\text{nominal}}$ denotes the nominal branching fraction result, $V_{\text{test}}$ denotes the branching fraction from a specific mass window, $\sigma_{\mathrm{V}\text{nominal}}$ and $\sigma_{\mathrm{V}\text{test}}$ are the statistical uncertainty of the nominal branching fraction and the uncertainty corresponding to $V_{\text{test}}$, respectively.
  If $\zeta<2$ for all mass windows, the difference in the measured branching fractions is considered to be due to statistical fluctuations, rendering the uncertainty negligible. Otherwise, the largest difference in the measured branching fraction, corresponding to the highest value of $\zeta$, is taken as the uncertainty.

The uncertainty of the fit range is found to be negligible using the Barlow test. The uncertainty arising from the signal shape is estimated by substituting the signal function derived from the simulated shape with a double Gaussian function. The difference in the fitted signal yield, amounting to 1.6\%, is taken as the systematic uncertainty.
 The uncertainty related to the background function is estimated by replacing the Chebyshev polynomial function with the 2-D $\Sigma^{0}\bar{\Sigma}^{0}$ sideband. The difference in the fitted signal yield, 0.6\%, is considered as the uncertainty. 

The uncertainty associated with the MC model is estimated by using the data-driven {\tt BODY3}~\cite{evtgen_a} generator to regenerate a signal MC sample of equivalent size as the nominal signal MC sample. The Dalitz plot of $M^{2}_{\Sigma^{0}\omega}$ versus $M^{2}_{\bar{\Sigma}^{0}\omega}$ is input into the {\tt BODY3} generator, with backgrounds subtracted using the $\omega$ sideband. The efficiency difference between the alternative and nominal MC samples, 7.0\%, is assigned as a systematic uncertainty.

The uncertainty due to the limited signal MC sample size is calculated as
$\sqrt{\frac{1-\varepsilon}{N \cdot \varepsilon}}=0.7\%
$, where $\varepsilon$ is the efficiency and $N$ is the total number of generated MC events.

   The uncertainties for $\Br\left(\Lambda\to p \pi^{-}/\bar{\Lambda}\to \bar{p} \pi^{+}\right)$ and $\Br(\omega \to \pi^{+} \pi^{-} \pi^{0})$ are sourced from the PDG~\cite{pdg2022}, which are 0.8\% for each decay. The uncertainty of $\Br(\pi^{0}\to\gamma\gamma)$ is considered negligible.

 The uncertainty associated with the total number of $\psi(3686)$ events is 0.5\%~\cite{psip_num_21}.

\begin {table}[tb]
\renewcommand\arraystretch{1.2}
{\caption {Relative systematic uncertainties for the branching fraction measurement of $\psi(3686)\to\Sigma^{0}\bar{\Sigma}^{0}\omega$.}
\label{uncertainty}}
\begin{tabular}{lc}
\hline
\hline

  Source  &         Uncertainty in (\%)       \\  \hline
        Tracking   & 2.0     \\
        Photon reconstruction     &  2.0   \\
       $\pi^{0}$ reconstruction    &  1.0  \\        
        $\Lambda$ $\bar{\Lambda}$ reconstruction & 1.8    \\        
        5C kinematic fit  &   1.1      \\
         $\Sigma^{0}$$\bar{\Sigma}^{0}$ mass window &   0.1    \\
Veto $\pi^{+}\pi^{-}J/\psi$& Negligible\\ Veto $\pi^{0}\pi^{0}J/\psi$& 4.1\\    
       Fit range     &     Negligible        \\       
       Signal shape       &     1.6        \\
       Background shape       &      0.6        \\ 
      Signal MC sample size&   0.7      \\
      Generator model   &   7.0   \\
        $\omega$ branching fraction&    0.8  \\
        $\Lambda$ $\bar{\Lambda}$ branching fraction&   1.6    \\
        $\pi^{0}$ branching fraction&    Negligible\\
       Total number of $\psi(3686)$ events&     0.5      \\
       \hline
       Total &    9.2     \\

\hline
\hline

\end{tabular}
\end{table}

	\section{Summary}\label{sec:summary}\vspace{-0.2cm}
 
   Using $(27.12\pm0.14)\times10^{8}$ $\psipp$ events collected by
   the BESIII detector in $2009$, $2012$ and $2021$, we report the first observation of the decay $\psi(3686)\to\Sigma^{0}\bar{\Sigma}^{0}$$\omega$ with a statistical significance of 8.9$\sigma$. The branching
   fraction is measured to be $(1.24 \pm 0.16_{\textrm{stat}} \pm 0.11_{\textrm{sys}})\times 10^{-5}$. This result is comparable to the previous measurement of its isospin partner process, $\psi(3686) \rightarrow\Sigma^{+}\bar{\Sigma}^{-}\omega$~\cite{SSphi}, which was found to be $(1.89 \pm 0.18_{\textrm{stat}} \pm 0.21_{\textrm{sys}})\times 10^{-5}$. We also investigate potential structures in the invariant mass distributions of $\Sigma^{0}\omega$, $\bar{\Sigma}^{0}\omega$ and $\Sigma^{0}\bar{\Sigma}^{0}$. A hint of a resonance in the $M_{\Sigma^{0}(\bar{\Sigma}^0)\omega}$ spectrum is observed with a significance of $2.5\sigma$.
   However, due to the limited statistics, we do not consider possible interference effects, and detailed information about the intermediate state, including its quantum numbers, cannot be determined.

 
\begin{center}
  \textbf{Acknowledgement}  
\end{center}

The BESIII Collaboration thanks the staff of BEPCII (https://cstr.cn/31109.02.BEPC) and the IHEP computing center for their strong support. This work is supported in part by National Key R\&D Program of China under Contracts Nos. 2020YFA0406300, 2020YFA0406400, 2023YFA1606000, 2023YFA1606704; National Natural Science Foundation of China (NSFC) under Contracts Nos. 11635010, 11735014, 11935015, 11935016, 11935018, 12025502, 12035009, 12035013, 12061131003, 12192260, 12192261, 12192262, 12192263, 12192264, 12192265, 12221005, 12225509, 12235017, 12361141819; the Chinese Academy of Sciences (CAS) Large-Scale Scientific Facility Program; the CAS Center for Excellence in Particle Physics (CCEPP); Joint Large-Scale Scientific Facility Funds of the NSFC and CAS under Contract No. U1832207; CAS under Contract No. YSBR-101; 100 Talents Program of CAS; The Institute of Nuclear and Particle Physics (INPAC) and Shanghai Key Laboratory for Particle Physics and Cosmology; Agencia Nacional de Investigación y Desarrollo de Chile (ANID), Chile under Contract No. ANID PIA/APOYO AFB230003; German Research Foundation DFG under Contract No. FOR5327; Istituto Nazionale di Fisica Nucleare, Italy; Knut and Alice Wallenberg Foundation under Contracts Nos. 2021.0174, 2021.0299; Ministry of Development of Turkey under Contract No. DPT2006K-120470; National Research Foundation of Korea under Contract No. NRF-2022R1A2C1092335; National Science and Technology fund of Mongolia; National Science Research and Innovation Fund (NSRF) via the Program Management Unit for Human Resources \& Institutional Development, Research and Innovation of Thailand under Contract No. B50G670107; Polish National Science Centre under Contract No. 2019/35/O/ST2/02907; Swedish Research Council under Contract No. 2019.04595; The Swedish Foundation for International Cooperation in Research and Higher Education under Contract No. CH2018-7756; U. S. Department of Energy under Contract No. DE-FG02-05ER41374




\begin{thebibliography}{0}%
\makeatletter
\providecommand \@ifxundefined [1]{%
 \@ifx{#1\undefined}
}%
\providecommand \@ifnum [1]{%
 \ifnum #1\expandafter \@firstoftwo
 \else \expandafter \@secondoftwo
 \fi
}%
\providecommand \@ifx [1]{%
 \ifx #1\expandafter \@firstoftwo
 \else \expandafter \@secondoftwo
 \fi
}%
\providecommand \natexlab [1]{#1}%
\providecommand \enquote  [1]{``#1''}%
\providecommand \bibnamefont  [1]{#1}%
\providecommand \bibfnamefont [1]{#1}%
\providecommand \citenamefont [1]{#1}%
\providecommand \href@noop [0]{\@secondoftwo}%
\providecommand \href [0]{\begingroup \@sanitize@url \@href}%
\providecommand \@href[1]{\@@startlink{#1}\@@href}%
\providecommand \@@href[1]{\endgroup#1\@@endlink}%
\providecommand \@sanitize@url [0]{\catcode `\\12\catcode `\$12\catcode `\&12\catcode `\#12\catcode `\^12\catcode `\_12\catcode `\%12\relax}%
\providecommand \@@startlink[1]{}%
\providecommand \@@endlink[0]{}%
\providecommand \url  [0]{\begingroup\@sanitize@url \@url }%
\providecommand \@url [1]{\endgroup\@href {#1}{\urlprefix }}%
\providecommand \urlprefix  [0]{URL }%
\providecommand \Eprint [0]{\href }%
\providecommand \doibase [0]{http://dx.doi.org/}%
\providecommand \selectlanguage [0]{\@gobble}%
\providecommand \bibinfo  [0]{\@secondoftwo}%
\providecommand \bibfield  [0]{\@secondoftwo}%
\providecommand \translation [1]{[#1]}%
\providecommand \BibitemOpen [0]{}%
\providecommand \bibitemStop [0]{}%
\providecommand \bibitemNoStop [0]{.\EOS\space}%
\providecommand \EOS [0]{\spacefactor3000\relax}%
\providecommand \BibitemShut  [1]{\csname bibitem#1\endcsname}%
\let\auto@bib@innerbib\@empty
\end{thebibliography}%


\begin{thebibliography}{**}

     \bibitem{BESIII_physics} D. M. Asner $et$ $al.$,    \href{https://doi.org/10.48550/arXiv.0809.1869}{Int. J. Mod. Phys. A \textbf{24}, S1 (2009).}

     \bibitem{QCD_charmonium2} N. Brambilla $et$ $al.$,    \href{https://doi.org/10.1140/epjc/s10052-010-1534-9}{Eur.~Phys.~J.C \textbf{71}, 1534 (2011).}

    \bibitem{baryons}V. Crede, \href{https://doi.org/10.1063/5.0009281} {AIP Conf. Proc. \textbf{2249}, 020003 (2020).}

    \bibitem{excited} A.~V. Sarantsev, M.~Matveev, V.~A.~Nikonov, A. V.
    Anisovich, U. Thoma, and E. Klempt, \href{https://doi.org/10.1140/epja/i2019-12880-5} {Eur. Phys. J. A \textbf{55}, 180 (2019).}

    \bibitem{hyperon}C. Fernández-Ramírez $et$ $al$., \href{https://doi.org/10.1103/PhysRevD.93.034029} {Phys. Rev. D \textbf{93}, 034029 (2016).}

        \bibitem{LLeta}M. Ablikim $et$ $al$. (BESIII Collaboration), \href{https://doi.org/10.1103/PhysRevD.106.072006} {Phys. Rev. D \textbf{106}, 072006 (2022).}

    \bibitem{LLomega}M. Ablikim $et$ $al$. (BESIII Collaboration), \href{https://doi.org/10.1103/PhysRevD.106.112011} {Phys. Rev. D \textbf{106}, 112011 (2022).}

    \bibitem{LLphi}M. Ablikim $et$ $al$. (BESIII Collaboration), \href{https://doi.org/10.1103/PhysRevD.104.052006} {Phys. Rev. D \textbf{104}, 052006 (2021).}

        \bibitem{psip_num_21} M.~Ablikim \textit{et al.} (BESIII Collaboration), \href{https://doi.org/10.1088/1674-1137/ad595b}{Chin. Phys. C \textbf{48}, 093001 (2024).}


     \bibitem{BES_design}M. Ablikim $et$ $al$. (BESIII Collaboration), \href{https://doi.org/10.1016/j.nima.2009.12.050} {Nucl. Instrum. Methods Phys. Res., Sect. A \textbf{614}, 345 (2010).}  
     
     \bibitem{Yu:IPAC2016-TUYA01} C.~H.~Yu {\it et al.}, \href{https://doi.org/10.18429/JACoW-IPAC2016-TUYA01}{Proceedings of IPAC2016, Busan, Korea, 2016.}

  \bibitem{Ablikim:2019hff}  M.~Ablikim {\it et al.} (BESIII Collaboration),
  \href{https://doi.org/10.1088/1674-1137/44/4/040001}{Chin. Phys. C {\bf 44}, 040001 (2020).}
  
  
  \bibitem{EventFilter}
  J.~W.~Zhang, {\it et al.},
  \href{https://doi.org/10.1007/s41605-022-00331-7}{Radiat. Detect. Technol. Methods {\bf 6}, 289–293 (2022).}  
  
   \bibitem{tof_a} X. Li \textit{et al.}, \href{https://doi.org/10.1007/s41605-022-00331-7}{Radiat. Detect. Technol. Meth. {\bf 1}, 13 (2017)}.


   \bibitem{tof_b} Y. X. Guo \textit{et al.}, \href{https://doi.org/10.1007/s41605-017-0012-4}{Radiat. Detect. Technol. Meth. {\bf 1}, 15 (2017)}.

   \bibitem{tof_c} P. Cao \textit{et al.}, \href{https://doi.org/10.1007/s41605-017-0014-2}{Nucl. Instrum. Methods Phys. Res., Sect. A {\bf 953}, 163053 (2020)}.

   \bibitem{geant4} S.~Agostinelli \textit{et al.} (GEANT4 Collaboration), \href{https://doi.org/10.1016/S0168-9002(03)01368-8}{Nucl. Instrum. Methods Phys. Res., Sect. A {\bf506}, 250 (2003).}

   \bibitem{detvis} K.~X.~Huang {\it et al.}, \href{https://doi.org/10.1007/s41365-022-01133-8}{Nucl.\ Sci.\ Tech. {\bf 33}, 142 (2022).}


   \bibitem{kkmc_a} S. Jadach, B. F. L. Ward, and Z. Was, \href{https://doi.org/10.1016/S0010-4655(00)00048-5}{Comput. Phys. Commun. {\bf 130}, 260 (2000)}.


    \bibitem{kkmc_b} S. Jadach, B. F. L. Ward, and Z. Was, \href{https://doi.org/10.1103/PhysRevD.63.113009}{Phys. Rev. D {\bf 63}, 113009 (2001)}.


    \bibitem{evtgen_a} R. G. Ping, \href{https://doi.org/10.1088/1674-1137/32/8/001}{Chin. Phys. C {\bf 32}, 599 (2008)}.


   \bibitem{evtgen_b} D. J. Lange, \href{https://doi.org/10.1016/S0168-9002(01)00089-4}{Nucl. Instrum. Methods Phys. Res., Sect. A {\bf 462}, 152 (2001)}.


      
        \bibitem{pdg2022}S.~Navas $et$ $al$. (Particle Data Group), \href{https://doi.org/10.1103/PhysRevD.110.030001}{Phys. Rev. D \textbf{110}, 030001 (2024)}.

    \bibitem{lundcharm_a} J. C. Chen, G. S. Huang, X. R. Qi, D. H. Zhang, and Y. S. Zhu, \href{https://doi.org/10.1103/PhysRevD.62.034003}{Phys. Rev. D {\bf 62}, 034003 (2000)}.


    \bibitem{lundcharm_b} R. L. Yang, R. G. Ping, and H. Chen, \href{https://doi.org/10.1088/0256-307X/31/6/061301}{Chin. Phys. Lett. {\bf31}, 061301 (2014)}.

     \bibitem{photos} E. Richter-Was, \href{https://doi.org/10.1016/0370-2693(93)90062-M} {Phys. Lett. B {\bf 303}, 163 (1993). }

       \bibitem{Sigma0_pair}M. Ablikim $et$ $al$. (BESIII Collaboration), \href{https://doi.org/10.1103/PhysRevLett.133.101902}{Phys. Rev. Lett. {\bf 133}, 101902 (2024)}.

       \bibitem{zhouxy_topoAna} X. Y. Zhou, S. X. Du, G. Li and C. P. Shen, \href{https://doi.org/10.1016/j.cpc.2020.107540}{Comput. Phys. Commun. {\bf 258}, 107540 (2021).}
       
     \bibitem{extend_likelihood} R. Barlow,  \href{https://doi.org/10.1016/0168-9002(90)91334-8} {Nucl. Instrum. Methods Phys. Res., Sect. A \textbf{297}, 496 (1990).} 

     \bibitem{keyspdf} K. S. Cranmer,  \href{https://doi.org/10.1016/S0010-4655(00)00243-5} {Comput. Phys. Commun. \textbf{136}, 3 (2001).} 

     \bibitem{LEE} O. Vitells and E. Gross,  \href{https://doi.org/10.1016/j.astropartphys.2011.08.005} {Astropart. Phys. \textbf{35}, 230 (2011).} 

       \bibitem{pion_tracking}M. Ablikim $et$ $al$. (BESIII Collaboration), \href{https://doi.org/10.1103/PhysRevD.87.012007}{Phys. Rev. D {\bf 87}, 012007 (2013)}.

       \bibitem{photon_rec}M. Ablikim $et$ $al$. (BESIII Collaboration), \href{https://doi.org/10.1103/PhysRevD.109.032006}{Phys. Rev. D {\bf 109}, 032006 (2024)}.

           \bibitem{SSphi}M. Ablikim $et$ $al$. (BESIII Collaboration), \href{https://doi.org/10.1103/PhysRevD.108.092011} {Phys. Rev. D \textbf{108}, 092011 (2023).}

       \bibitem{rec_lam(lamb)}M. Ablikim $et$ $al$. (BESIII Collaboration), \href{https://doi.org/10.1103/PhysRevD.108.112012 }{Phys. Rev. D {\bf 108}, 112012 (2023)}.

       \bibitem{5C_method}M. Ablikim $et$ $al$. (BESIII Collaboration), \href{https://doi.org/10.1103/PhysRevD.87.012002}{Phys. Rev. D {\bf 87}, 012002 (2013)}.

      \bibitem{Barlow_test}O. Behnke, K. Kroninger, G. Schott, and T. Schörner-Sadenius, ``How to Deal with Systematic Uncertainties," in {Data Analysis in high energy physics: A practical guide to statistical methods} (Wiley-VCH, Berlin, 2013).

	\end{thebibliography}
\end{document}